\documentclass[fleqn,10pt]{wlscirep}

\usepackage[normalem]{ulem}

\title{Super- and sub-radiance from two-dimensional resonant dipole-dipole interactions}

\author[1,*]{H. H. Jen}
\affil[1]{Institute of Physics, Academia Sinica, Taipei 11529, Taiwan}
\affil[*]{sappyjen@gmail.com}

\renewcommand{\r}{\mathbf{r}}

\renewcommand{\k}{\mathbf{k}}
\def\p{\mathbf{p}}
\def\q{\mathbf{q}}
\def\bea{\begin{eqnarray}}
\def\eea{\end{eqnarray}}
\def\ba{\begin{array}}
\def\ea{\end{array}}
\def\bdm{\begin{displaymath}}
\def\edm{\end{displaymath}}

\begin{abstract} 
We theoretically investigate the super- and sub-radiance from the resonant dipole-dipole interactions (RDDI) in a confined two-dimensional (2D) reservoir. The distinctive feature of 2D RDDI shows qualitatively and quantitatively different long-range behavior from RDDI in free space. We investigate the superradiant properties of the singly-excited symmetric state under this 2D RDDI. This state also allows subradiant decays in much longer distances than the transition wavelength, showing long-range atom-atom correlations. We further study the dynamics of the subradiant states which can be accessed by imprinting spatially dependent phases on the atomic arrays. Our results demonstrate rich opportunities in engineering light-matter interactions in a confined 2D reservoir, and hold promise in applications of quantum light storage and single-excitation state manipulations.
\end{abstract}
\begin{document}
\flushbottom
\maketitle
\section*{Introduction}

When light interacts with an ensemble of atoms, a spontaneous emission with an enhanced rate emerges due to strong atom-atom correlations induced by photons rescattering in the medium. This is so-called superradiance \cite{Dicke1954, Gross1982, Jen2012}, which can be manifested by resonant dipole-dipole interactions (RDDI) \cite{Stephen1964, Lehmberg1970, Jen2015}. The RDDI feature a long-range effect that connects every pair of the atoms, and thus enable the light-matter interacting system collectively coupled. This collective coupling is responsible for coherent light scattering \cite{Pellegrino2014, Jennewein2016, Jenkins2016, Araujo2016, Roof2016, Bromley2016, Zhu2016, Shahmoon2017}. On the other hand in a dense medium with RDDI, an emission with a reduced decay rate, the subradiance \cite{Sonnefraud2010, McGuyer2015, Guerin2016}, can also show up as an afterglow which follows the initialized superradiance in a cloud \cite{Guerin2016}. Many other proposals to prepare and manipulate the subradiant states can be found in free-standing atomic arrays \cite{Scully2015, Facchinetti2016, Jen2016_SR, Jen2016_SR2, Sutherland2016, Bettles2016, Garcia2017, Jen2017_MP, Jen2018_SR1, Jen2018_SR2, Bhatti2018}, atoms in a cavity \cite{Plankensteiner2017}, and metamolecules \cite{Jenkins2017}.

In addition to RDDI in a free space or three-dimensional (3D) reservoir, this near-resonantly driven dipole-dipole interaction can be also established in a confined one-dimensional (1D) atom-nanophotonic waveguide system \cite{Kien2005, Kien2008, Tudela2013, Kumlin2018, Chang2018}, which shows infinite-range couplings in sinusoidal forms and is observed recently in two atomic clouds above the optical nanofibers \cite{Solano2017}. By structuring 1D reservoir \cite{Scelle2013, Chen2014, Ramos2014}, the confined 1D system can further allow an effective uni-directional decay channel, which breaks the time reversal symmetry, and construct a chiral quantum optical network \cite{Pichler2015, Lodahl2017}. In such confined systems, the guided mode of light can be stored via electromagnetically induced transparency \cite{Sayrin2015}, and long-range quantum magnetism can be simulated with tunable spin-spin interactions \cite{Hung2016} where a pseudo spin is presented by two atomic internal states.

The RDDI involve the coherent and dissipative parts, which respectively determine the frequency shift and line width of the radiation. For the dissipative parts, Dicke's regime can be reached in short distance of $\xi\ll 1$ where $\xi$ represents a dimensionless scale of mutual separation. In opposite regime of long distance that $\xi\gg 1$ for both coherent and dissipative parts, 1D RDDI sustain the interaction strength as $\cos(\xi)$ and $\sin(\xi)$ respectively, in contrast to $\sim 1/\xi$ of 3D RDDI. Another stark contrast of 1D and 3D RDDI lies in the short distance asymptotics of the coherent parts, which are $\sim \xi$ and $\sim 1/\xi^3$ respectively. The divergence of 3D RDDI in the limit of $\xi\rightarrow 0$ indicates a breakdown of quantum optical treatment, which means a lack of complete and genuine description of induced dipoles in short distance. On the other hand in 1D confinement, $\xi$ in general can not be made infinitely small, while its interaction strength of $\sin(\xi)$ already diminishes as $\xi\ll 1$. Other than these well-known 3D and 1D RDDI, two-dimensional (2D) RDDI is less explored since it requires intricately designed 2D reservoir. Here we consider a setting of 2D optical lattices of atoms inside a planar cavity with perfect mirrors. This way, the atoms can only dissipate by spontaneous emissions into this 2D reservoir. Other potential settings for this confined 2D dissipative bath can be a 2D lattice of coupled ring resonators \cite{Mittal2018} or 2D arrays of superconducting artificial atoms.

In this article, we show the distinctive feature of 2D RDDI, where different atomic polarizations display significantly distinct long range behavior. We investigate the superradiant properties under a symmetric state of single photon excitation. This state also allows subradiant decays at some specific or much longer $\xi$, showing long-range atom-atom correlations. We further study the time dynamics of phase-imprinted subradiant states by applying spatially dependent phases on the 2D atomic arrays. The radiation properties of these potentially controllable single-excitation subradiant states highly depend on the 2D lattice structures. Thus, it allows engineering of light-matter interactions, and promises applications in quantum light storage and state manipulations.

\section*{Collective properties from RDDI in a confined two-dimensional reservoir}

\subsection*{RDDI in a confined two-dimensional reservoir}

We start by considering a confined two-dimensional (2D) reservoir, where 2D lattice array of $N$ two-level atoms are situated. We follow the general formalism of RDDI in 3D free space \cite{Lehmberg1970} in Appendix. From equation (\ref{J}), the 2D reservoir has a quantization area $A$, and we obtain the 2D RDDI of $J_{\mu,\nu}$ in polar coordinates,
\bea
J_{\mu,\nu}=\int_0^\infty\frac{q\bar g_q^2 A}{(2\pi)^2}  dq\int_0^{2\pi} d\theta [1-(\hat \q\cdot\hat\p)^2]e^{i\k_q\cdot\r_{\mu,\nu}}[\pi\delta(\omega_q-\omega_e)+\pi\delta(\omega_q+\omega_e)+i\mathcal{P}(\omega_e-\omega_q)^{-1}-i\mathcal{P}(\omega_q+\omega_e)^{-1}],
\eea
where $\r_{\mu,\nu}=\r_\mu-\r_\nu$, and $\hat\p$ denotes the excitation polarization. We calculate the real part of $J_{\mu,\nu}$ first, and $\hat\q$ in general has a polar angle $\theta$ to the $\hat y$ on the $\hat x-\hat y$ plane. Without loss of generality, we assume $\r_{\mu,\nu}$ along $\hat y$ and $\hat\p$ with a polar angle $\theta'$ to $\hat y$. We obtain
\bea
{\rm Re}[J_{\mu,\nu}]&=&\int_0^\infty\frac{q\bar g_q^2 A}{(2\pi)^2}  dq\int_0^{2\pi} d\theta[1-(\cos\theta\cos\theta'+\sin\theta\sin\theta')^2]e^{i\xi\cos\theta}\pi\delta(\omega_q-\omega_e),\nonumber\\
&=&\frac{\Gamma_{2D}}{2}\frac{1}{\pi}\int_0^{2\pi} d\theta[1-(\cos\theta\cos\theta'+\sin\theta\sin\theta')^2] e^{i\xi\cos\theta},
\eea 
where the intrinsic decay constant for the 2D reservoir is $\Gamma_{2D}$ $\equiv$ $2k_L|\partial_\omega q(\omega)|_{\omega=\omega_e}\bar g_{k_L}^2A/4$ with the coupling strength $\bar g_{k_L}$, the inverse group velocity $\partial_\omega q(\omega)$, and the quantization area $A$. The dimensionless atomic separation is $\xi\equiv k_L|\r_\mu-\r_\nu|$ with the near-resonant excitation wave number $k_L=\omega_e/c$. Integrating out the polar angles, we obtain
\bea
{\rm Re}[J_{\mu,\nu}]=\frac{\Gamma_{2D}}{2}2\left[J_0(\xi)-\frac{J_1(\xi)}{\xi}+(\hat\p\cdot\hat\r_{\mu,\nu})^2J_2(\xi)\right]
\equiv\frac{\Gamma_{2D}}{2}f(\xi),
\eea
where $\hat\r_{\mu,\nu}=(\r_\mu-\r_\nu)/|\r_\mu-\r_\nu|$. The above results can be derived from the following integrals,
\bea
&&\int_0^{2\pi}e^{ia\cos\theta}d\theta=2\pi J_0(|a|),\\
&&\int_0^{2\pi}\cos^2\theta e^{ia\cos\theta}d\theta=2\pi \left(\frac{J_1(a)}{a}-J_2(a)\right),\\
&&\int_0^{2\pi}\sin^2\theta e^{ia\cos\theta}d\theta=2\pi \frac{J_1(|a|)}{|a|},\\
&&\int_0^{2\pi}\sin\theta\cos\theta e^{ia\cos\theta}d\theta=0,
\eea
where $J_n(a)$ are the Bessel functions of the first kind. 

The ${\rm Re}[J_{\mu,\nu}]$ and ${\rm Im}[J_{\mu,\nu}]$ should satisfy the Kramers-Kronig relation, and therefore we derive $J_{\mu,\nu}$,
\bea
J_{\mu,\nu}&=&\frac{\Gamma_{2D}}{2}f(\xi)-\frac{i\mathcal{P}}{2\pi}\int_{0}^\infty d\omega(k_L|\partial_\omega q(\omega)|\bar g_q^2A/2) f(\omega|\r_\mu-\r_\nu|/c)\left(\frac{1}{\omega-\omega_e}+\frac{1}{\omega+\omega_e}\right),\\
&=&\frac{\Gamma_{2D}}{2}[f(\xi)+ig(\xi)],
\eea
where 
\bea
f(\xi)&\equiv& 2\left[J_0(\xi)-\frac{J_1(\xi)}{\xi}+(\hat\p\cdot\hat\r_{\mu,\nu})^2J_2(\xi)\right],\\ 
g(\xi)&\equiv& 2Y_0(\xi)-2\frac{Y_1(\xi)}{\xi}+2(\hat\p\cdot\hat\r_{\mu,\nu})^2Y_2(\xi)-\frac{4}{\pi\xi^2}[1-2(\hat\p\cdot\hat\r_{\mu,\nu})^2],
\eea
and $Y_n(\xi)$ are the Bessel functions of the second kind. The above $g(\xi)$ can be derived by using the following integrals,
\bea
&&\mathcal{P}\int_0^\infty da\frac{J_0(a)}{a\mp b}=-\frac{\pi}{2}[Y_0(b)\pm H_0(b)],\\
&&\mathcal{P}\int_0^\infty da\frac{J_1(a)}{a(a\mp b)}=-\frac{2+\pi b[Y_1(b)\pm H_1(b)]}{2b^2},\\
&&\mathcal{P}\int_0^\infty da\left(\frac{J_2(a)}{a- b}+\frac{J_2(a)}{a+ b}\right)=-\frac{4}{b^2}-\pi Y_2(b),
\eea
where $H_n(b)$ is the Struve function.

The dynamical equations for any atomic observables $Q\equiv\langle\hat Q\rangle$ tracing over the fields can be expressed in Lindblad forms (See Methods),
\bea
\dot Q(t) &=& \sum_{\nu\neq\mu}^N\sum_{\mu=1}^Ni{\rm Im}(J_{\mu,\nu})[\sigma^\dag_\mu\sigma_\nu,Q]+\mathcal{L}[Q],\\
\mathcal{L}[Q]&=&\sum_{\nu=1}^N\sum_{\mu=1}^N{\rm Re}(J_{\mu,\nu})\Big[\sigma_\mu^\dag Q\sigma_\nu-\frac{1}{2}(\sigma_\mu^\dag\sigma_\nu Q+Q\sigma_\mu^\dag\sigma_\nu)\Big],
\eea
where $\sigma_\mu^\dag\equiv|e\rangle\langle g|$ denotes the raising operator with $|g\rangle$ and $|e\rangle$ for the ground and excited states respectively. The coherent and dissipative coupling forms of $g(\xi)$ and $f(\xi)$ respectively denote the frequency shifts and decay rates between any pairs of the atoms. They, equivalently dispersion and absorption respectively, should satisfy the Kramers-Kr\"onig relation, as required of causality in a physical response function \cite{Zangwill2013}.

\begin{figure}[t]
\centering
\includegraphics[width=16cm,height=7.5cm]{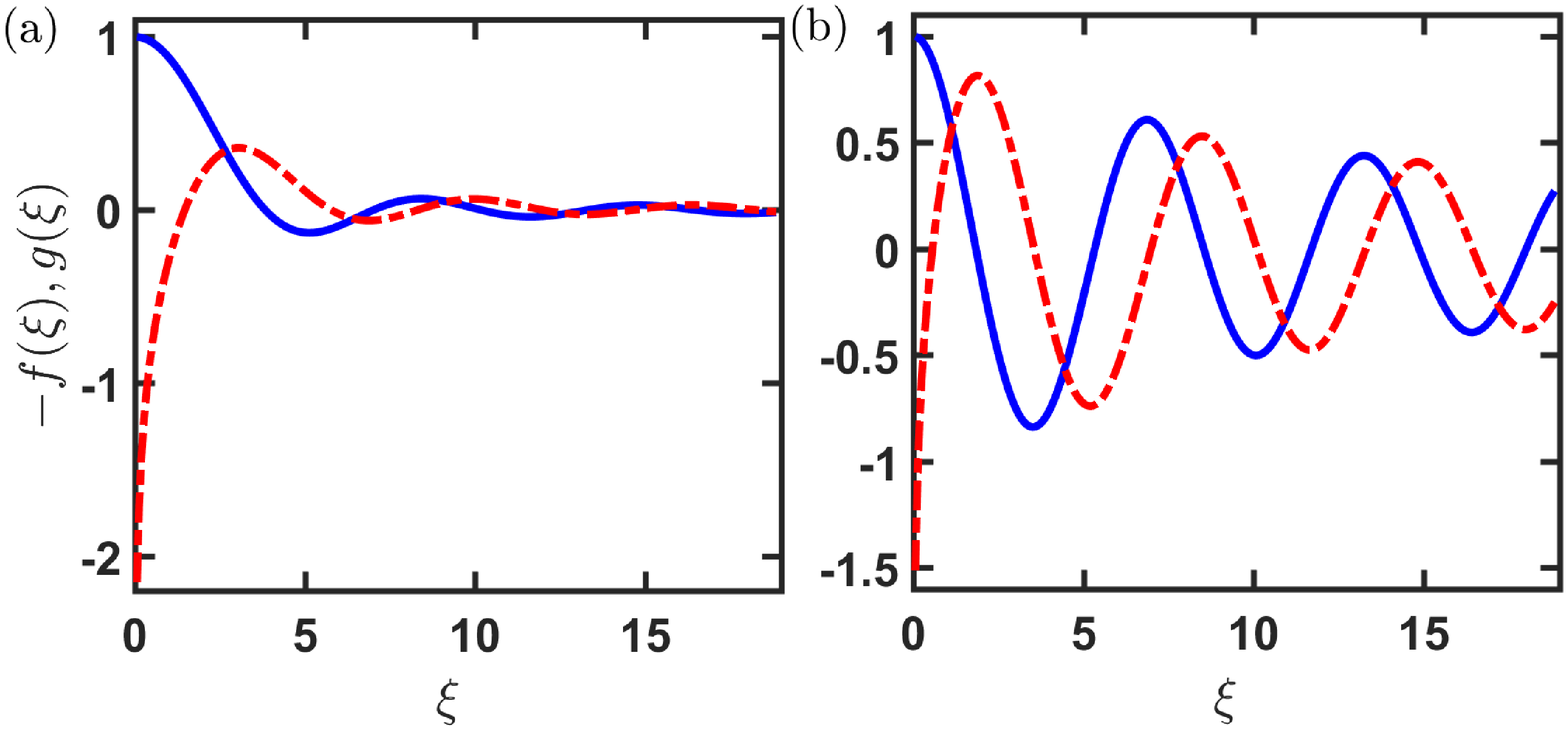}
\caption{Resonant dipole-dipole interactions in a 2D reservoir. The pairwise couplings strengths of collective decay rates [$-\Gamma_{2D}f(\xi)/2$, solid] and frequency shifts [$\Gamma_{2D}g(\xi)/2$, dash-dotted] are shown for light polarizations (a) parallel and (b) perpendicular to atomic separations $\hat\r_{\mu,\nu}$.}\label{fig1}
\end{figure}

In figure \ref{fig1}, we plot 2D RDDI for two orthogonal light polarizations with $\hat\p\parallel \hat\r_{\mu,\nu}$ and $\hat\p\perp\hat\r_{\mu,\nu}$ respectively. Both dissipative parts at small $\xi$ are similar and approach unity, which are within Dicke's superradiant regime. For the coherent parts in the same limit, the leading order of the asymptotics is $2\ln(\xi/2)/\pi$, which is non-analytic at $\xi=0$ but diverges much slower than $1/\xi^3$ in 3D RDDI. As shown in figure \ref{fig1}, the frequency shift is still in the order of $\Gamma_{2D}$ in as short as $\xi/(2\pi)=0.01$, where $|g(\xi)|\sim 2.1$ and $1.5$ respectively. This shows a prevailing effect of 2D RDDI on the radiations at such small scale of $\xi$, in huge contrast to the 3D case where divergent frequency shift forbids any atomic excitations. This promises a short-range and strongly interacting regime in the 2D RDDI, similar to the 1D case as its coherent parts $\sin\xi\approx 0$. For longer $\xi\gg 1$, $f(\xi)\rightarrow 1/\xi^{3/2}$ and $1/\sqrt{\xi}$ for respective polarization configurations, in contrast to the asymptotic forms of RDDI in free space, which are $1/\xi^2$ and $1/\xi$ respectively. This longer-range dependence of $1/\sqrt{\xi}$ is evident in figure \ref{fig1}(b), which can be seen as a crossover from 3D to 1D RDDI that eventually lead to infinite-range couplings. This length scaling in this particular polarization configuration can be reinterpreted by $\xi^{-(d-1)/2}$ where $d$ represents the dimension of reservoir from which RDDI emerge. As a consequence, the $f(\xi)$ in the case of $\hat\p\perp\hat\r_{\mu,\nu}$ weakens less rapidly over distances, which can still maintain a significant strength of $|f(\xi)|/f(0)\sim 50\%$ at $\xi\gtrsim 10$. This will make a significant effect on super- and subradiant properties, which are unique from the results in 1D and 3D reservoirs. 

\subsection*{Collective super- and subradiant couplings}

\begin{figure}[t]
\centering
\includegraphics[width=16cm,height=7.5cm]{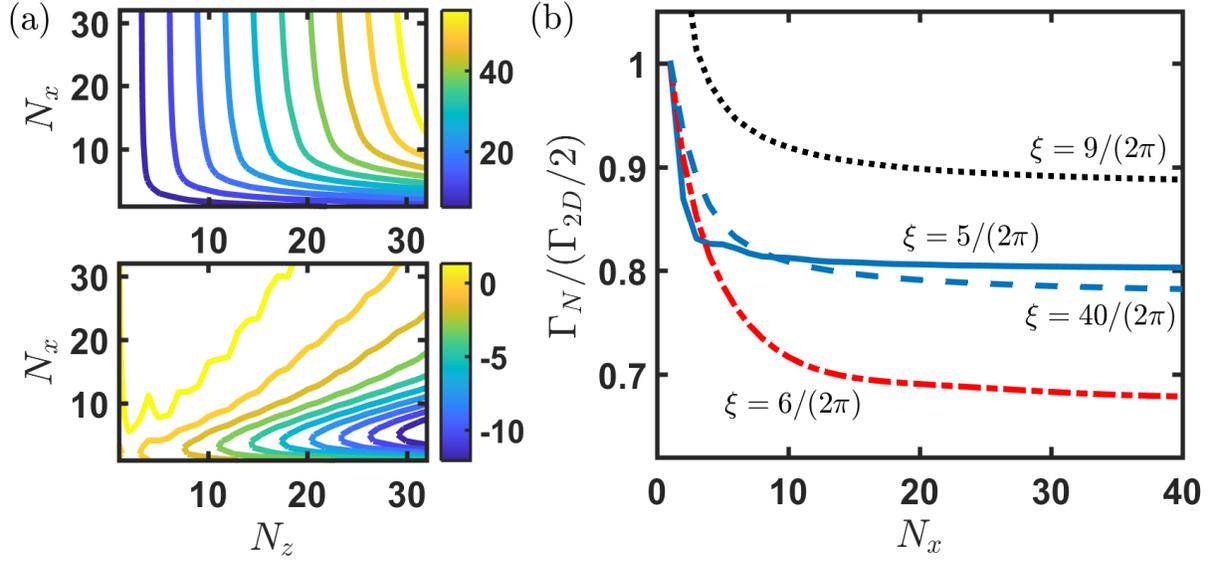}
\caption{Super- and subradiant decay constants and frequency shifts. (a) Superradiant decay constants $\Gamma_N$ and frequency shifts $\Delta_N$ are shown in the upper and lower panels respectively at $\xi=1$ with $\hat\p\parallel \hat x$. (b) Some selective $\xi$'s show subradiant decay behaviors as a dependence of $N_x$ for $N_z=1$ (solid and dash-dotted), $2$ (dotted), and $3$ (dashed), in the same configuration of $\hat\p$.}\label{fig2}
\end{figure}

In the following, we investigate the collective decay constants in a 2D lattice with 2D RDDI. We consider single photon interacting with an equidistant atomic array, and on absorption the atoms can be excited to the symmetric state,
\bea
|\Psi\rangle=\frac{1}{\sqrt{N}}\sum_{\mu=1}^Ne^{i\k_L\cdot\r_\mu}\sigma_\mu^\dag|0\rangle,\label{sym}
\eea 
where $e^{i\k_L\cdot\r_\mu}$ is the traveling phase carried by the photon. From the pairwise couplings under the symmetric state, that is $\langle\Psi|J_{\mu,\nu}\sigma^\dag_\mu\sigma_\nu|\Psi\rangle$, we obtain the cooperative superradiant constants and associated frequency shift respectively,
\bea
\Gamma_N=\frac{1}{N}\sum_{\nu=1}^N\sum_{\mu=1}^N e^{-i\k_L\cdot(\r_\mu-\r_\nu)}{\rm Re}[J_{\mu,\nu}],\\
\Delta_N=\frac{1}{N}\sum_{\nu\neq\mu}^N\sum_{\mu=1}^N e^{-i\k_L\cdot(\r_\mu-\r_\nu)}{\rm Im}[J_{\mu,\nu}],
\eea
from which the radiation intensity of spontaneously emitted photon can be described by a simple form of $\exp(-\Gamma_Nt+i\Delta_Nt)$. In figure \ref{fig2}(a), we show the superradiant properties of the symmetric state in a 2D $N_x\times N_z$ array with $\k_L$ along $\hat z$. In Dicke's limit where $\xi\lesssim 1$, we expect of similar $\Gamma_N$ from 2D or 3D RDDI in the same lattice configurations. $\Gamma_N$ saturates quite fast as $N_x\gg N_z$, showing an independence of the number of atoms as $N_x$ increases in the direction perpendicular to the light excitation. On the contrary, two contrasting dependences of $N_z$ can be located at $N_x\ll 10$ and $N_x\gtrsim 10$, which are $\sim N_z^{0.65}$ and $N_z^{0.97}$ respectively for $N_x=2$ and $30$. This shows a suppressed scaling in a needle-like 2D lattice compared to the square structure. Similar distinguishing features are also present in $\Delta_N$, where the needle-like structure allows significant red shifts, whereas for $N_z\lesssim N_x$, blue shifts emerge instead. In the region of $N_z<N_x$, we have a relatively broad and less varying dependence of lattice structures. 

More interesting decay behavior of 2D RDDI results from the oscillatory negative couplings in the case of $\hat\p\perp\hat\r_{\mu,\nu}$ in figure \ref{fig1}(b). As shown in figure \ref{fig2}(b), the subradiant decay can be supported at some selective $\xi$'s in an optically-thin lattice structure. This even sustains in longer distance, for example of $\xi=40$ in the plot. If we put this 2D lattice mediating 3D RDDI in free space, the subradiance under the symmetric states becomes less significant ($\sim 15\%$ more of the $\Gamma_N$ at $N_x\sim 40$), and thus 2D RDDI comparing the 3D case specifically show a notable long-range effect, resembling the infinite-range sinusoidal forms of 1D RDDI.

\section*{Phase-imprinted subradiant states}

\begin{figure}[t]
\centering
\includegraphics[width=16cm,height=7.5cm]{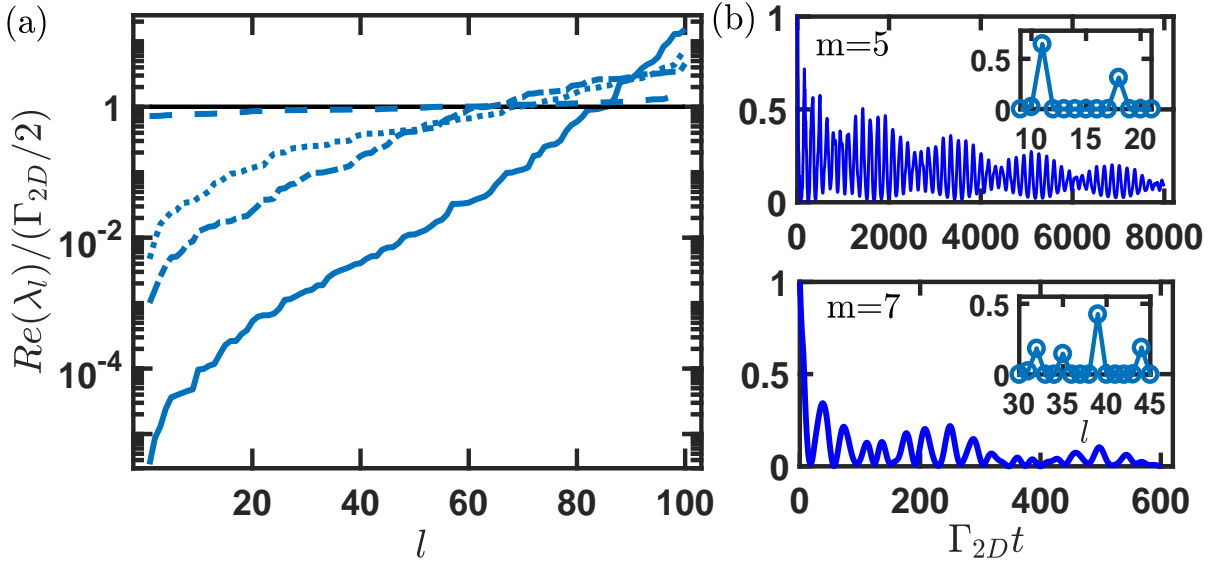}
\caption{Eigen-decay rates and time evolutions of subradiant states for $10\times 10$ array. (a) The eigen-decay constants can be obtained by the real parts of the eigenvalues $\lambda_l$ at $\xi=1$ (solid), $5$ (dash-dotted), $10$ (dotted). As a comparison, we show the results for the same 2D lattice configuration from 3D RDDI at $\xi=10$ (dashed), and the horizontal line guides the eye for a natural decay constant. (b) Time dynamics of some selective subradiant states of $m=5$ and $7$ at $\xi=1$, with most notable state weightings $|w_l(m)|^2$ on the eigenmodes in the respective insets. Clear beatings in the radiation can be seen in both plots due to finite Im($\lambda_l$).}\label{fig3}
\end{figure}

Next, we further study the subradiance from 2D RDDI, which can be enabled by imprinting linearly increasing phases on the atomic arrays \cite{Jen2016_SR, Jen2016_SR2} or via a side excitation with a $\pi$ phase shift in the sub-ensembles \cite{Scully2015}. This spatially varying phase can be imprinted by applying pulsed gradient magnetic or electric fields, or directly using light which carries orbital angular momentum in atomic ring structures \cite{Jen2018_SR1, Jen2018_SR2}. We first construct a complete Hilbert space of single excitation, which reads
\bea
|\Phi_m\rangle=\frac{1}{\sqrt{N}}\sum_{\mu=1}^Ne^{i\k_L\cdot\r_\mu}e^{i2m\pi(\mu-1)/N}\sigma_\mu^\dag|0\rangle,\label{DM}
\eea 
where $m\in[1,N]$. The above set should be orthonormal, where the inner products of them is $\langle\Phi_m|\Phi_n\rangle=N^{-1}\sum_{\mu=1}^N$ $e^{\frac{i2\pi(\mu-1)(m-n)}{N}}$ $=$ $\delta_{m,n}$, satisfied by De Moivre formula. The Hilbert space of equation (\ref{DM}) includes both super- and subradiant states, which has also been applied in forward- and backward-propagating eigenstates to reveal the emergent universal dynamics in a 1D nanophotonic system \cite{Kumlin2018}. Note that in general there are infinite ways to construct singly-excited Hilbert space, and therefore equation (\ref{DM}) is not unique to the setting of $N$ atoms interacting with single photon. The states of equation (\ref{DM}) can be prepared collectively where all atoms are excited uniformly, and allow studies on super- and sub-radiance systematically by varying the imprinted phases. Though these states can be controlled dynamically, their fidelities may suffer from an inefficient phase-imprinting protocol using pulsed lasers or limitation of large gradient magnetic fields \cite{Jen2016_SR}. Nevertheless, the phase imprinting construction allows a controllable way to manipulate these orthonormal states collectively, and thus the super- and subradiant properties correspondingly. 

Since 2D RDDI involve a long-range functional form, it is not possible to write down the analytical eigenstates in general. Therefore, we numerically derive the eigenbases, and the time evolutions of $|\Phi_m(t)\rangle$ can be obtained by solving the Schr\"{o}dinger equations, $\partial|\Phi(t)\rangle/\partial t$ $=$ $-J|\Phi(t)\rangle$, where the matrix elements of $J$ consists of the couplings $J^*_{\mu,\nu}$, and $|\Phi(t)\rangle=\sum_\mu b_\mu(t)\sigma_\mu^\dag|0\rangle$. By diagonalizing $J$, we obtain $\vec b(t)$ $=$ $Se^{\vec\lambda t}S^{-1}\vec b(t=0)$, where $S$ and $\vec\lambda$ are the eigenbases and eigenvalues respectively. For some initially prepared state $|\Phi(t=0)\rangle$ $=$ $|\Phi_m\rangle$, we obtain its time evolution as $A_m(t)\equiv\sum_l w_l(m)e^{\lambda_l t}$, where $w_l(m)$ $=$ $(h^\dag S)\cdot(S^{-1}h)$ leads to the weightings $|w_l(m)|^2$ of $|\Phi_m\rangle$ on respective $l$th eigenmodes, with a column vector $h$ consisting of the imprinted phases $N^{-1/2}e^{i\k\cdot\r_\mu+i2m\pi(\mu-1)/N}$.

In figure \ref{fig3}(a), we show the distributions of the eigenmodes in an ascending order in a 2D square lattice. When $\xi\lesssim 5$ or the mutual distance is less than the resonant wavelength, the 2D system allows significant super- and subradiant eigen-decay constants, as expected and similar to the results from 3D RDDI in a strongly interacting regime. By contrast, as $\xi$ extends further, 2D RDDI still permit the lowest decay rate below $10^{-2}\Gamma_{2D}$ in figure \ref{fig3}(a), indicating of long-range atom-atom correlations. As a comparison, in the same 2D lattice configuration but from a 3D reservoir, the eigenmodes show a level dependence and have reached the noninteracting regime. In figure \ref{fig3}(b), as an example, we further show the radiation intensity $|A_m(t)|^2$ of two subradiant states assuming they are initially created. Increasing $m$ means larger gradient fields required to prepare these states. For the selective state of $m=5$, we see quite a slow subradiant decay with two beating frequencies as time evolves. This originates from three dominating eigenmodes as shown in the weightings of the inset (less obvious for $|w_{10}(m)|\sim 0.15$), where two most significant modes of $l=11$ and $18$ occupy relatively small eigen-decay constants of $\sim 10^{-4}$ and $\sim 4\times 10^{-4}\Gamma_{2D}$ respectively. The beating frequencies can be determined by the differences of ${\rm Im}(\lambda_l)$ in the respective modes. Other example of $m=7$ state in the lower plot of figure \ref{fig3}(b) also shows the subradiant decay behavior with four significant weightings on the eigenmodes instead. The lifetime of the oscillatory waveform can be approximately decided by the most significant mode of $l=39$ which has an eigen-decay rate $\sim 4\times 10^{-3}\Gamma_{2D}$. In addition to the fast oscillating ripples in the radiation pattern in both figures \ref{fig3}(a) and \ref{fig3}(b), a slowly-varying envelope also appears and extends to long time scales, implying the dominance of the subradiant modes.    

\begin{figure}[t]
\centering
\includegraphics[width=16cm,height=7.5cm]{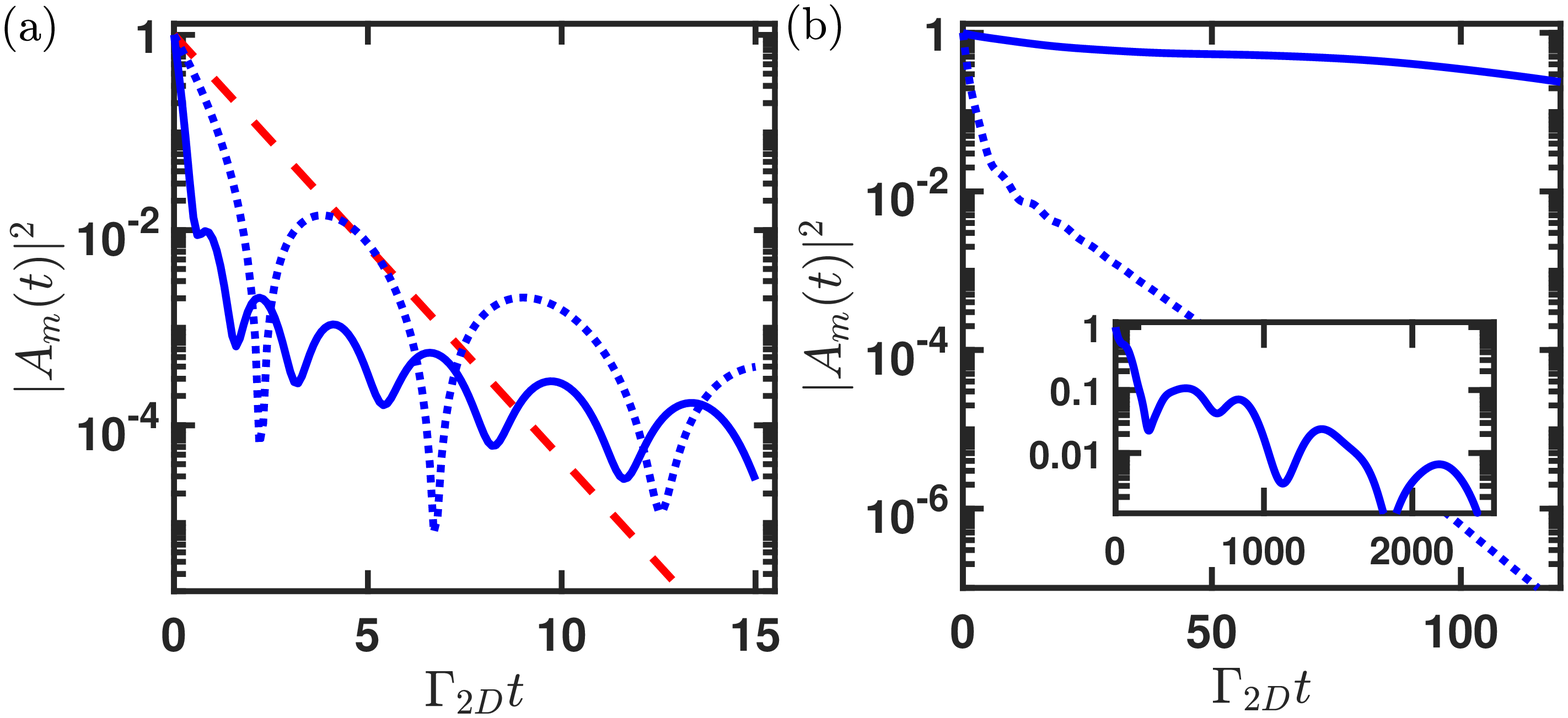}
\caption{Contrasting time evolutions at $\xi=5$ for a $N_x\times N_z=4\times 20$ array. (a) Superradiant state dynamics of $m=0$ with $\k_L$ along $\hat z$ (solid) and $\hat x$ (dotted) respectively, comparing the natural decay of $e^{-\Gamma_{2D}t}$ (dashed). (b) Subradiant state evolutions of $m=10$ with $\k_L$ along $\hat z$ (solid) and $\hat x$ (dotted) respectively. The inset shows the long time behavior for the subradiant state in the case of $\k_L$ along $\hat z$, which shows an early drop with an oscillatory subradiance afterward, similar to (a).}\label{fig4}
\end{figure}

Finally, we study a striped 2D lattice structure for two orthogonal light excitations in figure \ref{fig4}. For both super- and subradiant states in the example of 2D lattice with $N_z\gg N_x$, the decay behavior can be approximately separated into two time scales, an early fast drop and late subradiant decay, which also manifests in a dense 3D cloud \cite{Guerin2016}. In figure \ref{fig4}(a), the superradiant state of $m=0$ becomes exactly the symmetric state of equation (\ref{sym}), where $\Gamma_N$ governs the decay behavior in the beginning when $|A_m(t)|^2$ $\gtrsim$ $0.01$. The later oscillatory subradiance indicates of multiple though less occupied subradiant modes. Comparing the lifetime determined when its initial probability drops to $e^{-1}$ in the early stage, an optically-thick striped lattice in the case of $\k_L\parallel\hat z$ shows an enhanced decay rate by only a factor of $\sim 4$ over the case of $\k_L\parallel\hat x$. On the other hand for the subradiant states in figure \ref{fig4}(b) with a finite phase imprinting, the contrasting reduction factor of the decay rates becomes $\sim 100$ in the optically-thick configuration. This magnifying factor in the subradiant time scale suggests a potential photon routing relying on the 2D lattice mediating 2D RDDI, where light going through an optically-thick direction delays and almost stops within the time $\sim 100\Gamma_{2D}^{-1}$. Furthermore, potential chiral implementations using the phase-imprinted many-body states can be feasible in various atomic systems, for example cavity-optomechanical circuits \cite{Fang2017}, 2D coupled ring resonators \cite{Mittal2018}, or superconducting qubits and quantum dots in the photonic waveguides \cite{Barik2018} under an effectively emulated 2D reservoir.

\section*{Conclusion}

In conclusion, we have derived the explicit form of the RDDI from a confined two-dimensional reservoir. We demonstrate distinctive characteristics of 2D RDDI, which allows subradiance under a singly-excited symmetric state more significantly than the 3D case. This indicates long-range atom-atom correlations which are different from the induced RDDI in either 1D or 3D reservoirs. By imprinting spatially dependent phases on the 2D atomic arrays, we propose to prepare single-excitation subradiant states in a potentially deterministic and controllable way. Our results put forward potential applications in manipulating quantum information and preparations of many-body subradiant states in a 2D reservoir.


\section*{Methods}
\subsection*{General formalism for resonant dipole-dipole interaction in a three-dimensional reservoir}

Here we review the general formalism of resonant dipole-dipole interaction (RDDI) \cite{Stephen1964, Lehmberg1970} in a free space of three-dimensional (3D) reservoir. The RDDI originates from the common quantized light fields rescattering multiple times in the dissipation process. This collective dipole-dipole interaction in an ensemble of two-level quantum emitters is responsible for cooperative spontaneous emissions, so-called superradiance \cite{Dicke1954, Gross1982} and subradiance, and collective frequency shift \cite{Friedberg1973, Scully2009}. Only recently that significantly small collective frequency shift can be observed in some versatile atomic systems, including the embedded atoms in the planar cavity \cite{Rohlsberger2010}, a vapor cell \cite{Keaveney2012}, an ionic system \cite{Meir2014}, and cold atoms \cite{Pellegrino2014}.   

The spontaneous decay behavior in a system of $N$ two-level quantum emitters, with $|g\rangle$ and $|e\rangle$ for the ground and excited states respectively, can be described by a 3D reservoir of quantized bosonic light fields interacting with the medium. With a dipole approximation, the Hamiltonian reads \cite{Lehmberg1970},
\bea
H = \sum_{\mu=1}^N \hbar\omega_e\hat\sigma_\mu^\dag\hat\sigma_\mu-\sum_{\mu=1}^N\sum_q g_q (e^{i\k_q\cdot\r_\mu-i\omega_q t}\hat a_q+e^{-i\k_q\cdot\r_\mu+i\omega_q t}\hat a_q^\dag) (\hat \sigma_\mu +\hat\sigma_\mu^\dag),
\eea
where the atomic raising operator is $\hat\sigma_\mu^\dag\equiv|e\rangle_\mu\langle g|$ with $\hat\sigma_\mu=(\hat\sigma_\mu^\dag)^\dag$, and quantized fields $\hat a_q$ should satisfy the bosonic commutation relations $[\hat a_q,\hat a_{q'}^\dag]=\delta_{q,q'}$. The coupling constant $g_q\equiv d/\hbar\sqrt{\hbar\omega_q/(2\epsilon_0V)}(\vec\epsilon_q\cdot\hat d)$ involves a dipole moment $d$ with its unit direction $\hat d$, two possible polarizations of the fields $\vec\epsilon_q$ with the modes $q$, and a quantization volume $V$. The above Hamiltonian involves the non-rotating wave terms which are necessary for a complete description of the frequency shift (dispersion) of the RDDI in the dissipation. Therefore, the dispersion and absorption of RDDI should satisfy the Kramers-Kronig relation. 

Following the derivations in Ref. \cite{Lehmberg1970}, we continue to formulate a Heisenberg equation for an atomic operator $\hat Q$, that is $d\hat Q/dt=i[H,\hat Q]$ (let $\hbar=1$). We obtain 
\bea
\frac{d\hat Q}{dt}=i\omega_e\sum_\mu[\hat\sigma_\mu^\dag\hat\sigma_\mu,\hat Q]-i\sum_\mu\sum_q g_q\{e^{i\k_q\cdot\r_\mu}[\hat\sigma_\mu+\hat\sigma_\mu^\dag,\hat Q]\hat a_q(t)-e^{-i\k_q\cdot\r_\mu}\hat a_q^\dag(t)[\hat Q,\hat\sigma_\mu+\hat\sigma_\mu^\dag]\}.
\eea
In the above, $\hat a_q$ can be further substituted by solving $d\hat a_q/dt=i[H,\hat a_q]$, which is
\bea
\hat a_q(t)=\hat a_q(0)e^{-i\omega_q t}+i\sum_\mu g_qe^{-i\k_q\cdot\r_\mu}\int_0^t dt' [\hat\sigma_\mu(t')+\hat\sigma_\mu^\dag(t')]e^{-i\omega_q(t-t')}.
\eea
With the Born-Markov approximation of $\omega_e t\gg 1$ and $t\gg (r_{\mu\nu})_{max}/c$ ($r_{\mu\nu}\equiv |\r_\mu-\r_\nu|$), we obtain the dynamical equation of $Q\equiv\langle\hat Q\rangle_0$ in Lindblad forms by considering the vacuum initial bosonic fields $\langle\rangle_0$,
\bea
\dot{Q}(t)&=&\sum_{\mu\neq\nu}i\Omega_{\mu,\nu}[\sigma_\mu^\dag\sigma_\nu,Q]+\mathcal{L}(Q),\\
\mathcal{L}(Q)&=&\sum_{\mu,\nu}\gamma_{\mu,\nu}\left[\sigma_\mu^\dag Q\sigma_\nu-\frac{1}{2}(\sigma_\mu^\dag\sigma_\nu Q+Q\sigma_\mu^\dag\sigma_\nu)\right].
\eea 
The $\Omega_{\mu,\nu}$ and $\gamma_{\mu,\nu}$ describe the collective frequency shifts and decay rates respectively. These represent the coherent and dissipative parts of the pairwise couplings, $J_{\mu,\nu}\equiv(\gamma_{\mu,\nu}+i2\Omega_{\mu,\nu})/2$, which are defined as
\bea
J_{\mu,\nu}&=&\sum_q |g_q|^2\int_0^\infty dt' e^{i\k_q\cdot(\r_\mu-\r_\nu)}[e^{i(\omega_e-\omega_q)t'}+e^{-i(\omega_e+\omega_q)t'}], \nonumber\\
&=&\sum_q |g_q|^2\int_0^\infty dt' e^{i\k_q\cdot(\r_\mu-\r_\nu)}[\pi\delta(\omega_q-\omega_e)+\pi\delta(\omega_q+\omega_e)+i\mathcal{P}(\omega_e-\omega_q)^{-1}-i\mathcal{P}(\omega_q+\omega_e)^{-1}],\label{J}
\eea
where $\mathcal{P}$ is the principal value of the integral.

For a 3D reservoir, we consider continuous limits of modes $\sum_q\rightarrow\sum_{\vec\epsilon_q}\int_{-\infty}^\infty\frac{V}{(2\pi)^3}d^3q$ with two possible field polarizations $\vec\epsilon_q$. In spherical coordinates, we show the main results of $J_{\mu,\nu}$ in free space \cite{Lehmberg1970},
\bea
\gamma_{\mu,\nu}(\xi)&\equiv&\oint d\Omega_q[1-(\hat\q\cdot\hat\p)^2]\int_0^\infty dq q^2 \bar g_q^2\frac{V}{(2\pi)^3}[\pi\delta(\omega_q-\omega_e)+\pi\delta(\omega_q+\omega_e)],\nonumber\\
&=&\frac{3\Gamma}{2}\bigg\{\left[1-(\hat\p\cdot\hat{r}_{\mu\nu})^2\right]\frac{\sin\xi}{\xi}
+\left[1-3(\hat\p\cdot\hat{r}_{\mu\nu})^2\right]\left(\frac{\cos\xi}{\xi^2}-\frac{\sin\xi}{\xi^3}\right)\bigg\},\label{F}\\
\Omega_{\mu,\nu}(\xi)&\equiv&-\oint d\Omega_q[1-(\hat\q\cdot\hat\p)^2]\int_0^\infty dq q^2 \bar g_q^2\frac{V}{(2\pi)^3}[i\mathcal{P}(\omega_q-\omega_e)^{-1}+i\mathcal{P}(\omega_q+\omega_e)^{-1}],\nonumber\\
&=&\frac{3\Gamma}{4}\bigg\{-\Big[1-(\hat\p\cdot\hat{r}_{\mu\nu})^2\Big]\frac{\cos\xi}{\xi}
+\Big[1-3(\hat\p\cdot\hat{r}_{\mu\nu})^2\Big]
\left(\frac{\sin\xi}{\xi^2}+\frac{\cos\xi}{\xi^3}\right)\bigg\}\label{G}, 
\eea
where $d\Omega_q$ denotes an integration of a solid angle, $\bar g_q^2$ $\equiv$ $(d/\hbar)^2[\hbar\omega_q/(2\epsilon_0V)]$, $\hat\p$ parallels the excitation field polarization, the natural decay constant $\Gamma=d^2\omega_e^3/(3\pi\hbar\epsilon_0c^3)$, and dimensionless $\xi\equiv k_L|\r_\mu-\r_\nu|$ with $k_L=\omega_e/c$. As $\xi\rightarrow 0$, Dicke's regime is reached where $\gamma_{\mu,\nu}\rightarrow \Gamma$, while $\Omega_{\mu,\nu}$ goes to infinity. This divergence shows the inapplicability of quantum optical treatment in RDDI or in other words, it simply forbids any atomic excitations by external fields.


\section*{Acknowledgments}
We acknowledge the support from the Ministry of Science and Technology (MOST), Taiwan, under the Grant No. MOST-106-2112-M-001-005-MY3 and 107-2811-M-001-1524. We thank Y.-C. Chen, G.-D. Lin, and M.-S. Chang for insightful discussions, and are also grateful for NCTS ECP1 (Experimental Collaboration Program).

\section*{Author contributions statement}

H. H. Jen conducted the derivations and numerical simulations, analyzed the results, and wrote the manuscript.

\section*{Additional information}

\textbf{Competing financial interests:} The authors declare that they have no competing interests. 

\noindent\textbf{Publisher's note:} Springer Nature remains neutral with regard to jurisdictional claims in published maps and institutional affiliations.

\end{document}